\documentclass{ws-p8-50x6-00}

\begin{document}

\title{Space, Time, and Matter: Cosmological Parameters 2001}

\author{Lawrence M. Krauss}

\address{Departments of Physics and Astronomy \\
Case Western Reserve University,
10900 Euclid Ave. 
Cleveland, OH~~44106-7079}


\maketitle

\abstracts
{Over the past three years, our confidence in the inferred values of
cosmological parameters has increased dramatically, confirming
that the flat matter dominated Universe that dominated
cosmological model building for the past 20 years does not correspond to
the Universe in which we live.   I review recent developments here and
quote best fit current values for fundamental cosmological parameters.
\\ (Invited Review Lecture: Third International Conference on the 
Identification of Dark Matter, York, England, Sept 2000)}

\section{Introduction:}

The last decade has witnessed several remarkable transformations 
in our knowledge of the current state of our universe, including the
value of fundamental cosmological parameters.   
The good news is that the actual values we seem to be converging on are
ridiculous.  Thus, our increased empirical knowledge has gone hand in
hand with increased theoretical confusion.   For theorists and observers
alike, nothing could be more exciting.

In order to make some attempt to have this review appear less 
boring than it actually is, I have decided to group the cosmological
parameters in terms of three general themes, harking back to the
famous book of Hermann Weyl, the title of which nicely encompasses the
business of modern cosmology.

\section{Space, The Final Frontier:}

\subsection{Expansion:}  Probably the most important characteristic of
the  space in which we live is that it is expanding.  The expansion rate,
given by the Hubble Constant, sets the overall scale for most other
observables in cosmology.   Thus it is of vital importance to pin down
its value if we  hope to seriously constrain other cosmological
parameters..  

Fortunately, over the past five years tremendous strides have been 
made in our empirical knowledge of the Hubble constant.   I briefly
review recent developments and prospects for the future here.

\vskip 0.2in
\noindent{\bf \it HST-KEY Project:}

This is the largest scale endeavor carried out over the past decade 
with a goal of achieving a 10 $\%$ absolute uncertainty in the Hubble
constant.   The goal of the project has been to use Cepheid luminosity
distances to 25 different galaxies located within 25 Megaparsecs in order
to calibrate a variety of secondary distance indicators, which in turn
can be used to determine the distance to far further objects of known
redshift.  This in principle allows a measurement of the
distance-redshift relation and thus the Hubble constant on scales where
local peculiar velocities are insignificant.  The four distance
indicators so constrained are:  (1) the Tully Fisher relation,
appropriate for spirals, (2) the Fundamental plane, appropriate for
ellipticals, (3) surface brightness fluctuations, and (4) Supernova Type
1a distance measures.  

The HST-Key project has recently reported Hubble constant measurements 
for each of these methods, which I present below \cite{hst}.  While I
shall adopt these as quoted, it is worth pointing out that some critics
of this analysis have stressed that this involves utilizing data obtained
by other groups, who themselves sometimes report different values of the
Hubble constant for the same data sets.

$$ H_O^{TF} = 71 \pm 4 \pm 7  $$
$$ H_O^{FP} = 78 \pm 8 \pm 10  $$
$$ H_O^{SBF} = 69 \pm 4 \pm 6  $$
$$ H_O^{SN1a} = 68 \pm 2 \pm 5  $$
$$ H_O^{WA} =71 \pm 6 \ km s^{-1} Mpc ^{-1}  ( 1 \sigma)  $$

In the weighted average quoted above, the dominant contribution to 
the $9\%$ one sigma error comes from an overall uncertainty in the
distance to the Large Magellanic Cloud.   If the Cepheid Metallicity 
were shifted within its allowed  4$\%$ uncertainty range, the best fit
mean value for the Hubble Constant from the HST-Key project would shift
downard to  $68 \pm 6$.

\vskip 0.2in
\noindent{\bf \it S-Z Effect:}

The Sunyaev-Zeldovich effect results from a shift in the spectrum of 
the Cosmic Microwave Background radiation due to scattering of the
radiation by electgrons as the radiation  passes through intervening
galaxy clusters on the way to our receivers on Earth.  Because the
electron temperature in Clusters exceeds that in the CMB, the radiation
is systematically shifted to higher frequencies, producing a deficit in
the intensity below some characteristic frequency, and an excess above
it.  The amplitude of the effect depends upon the Thompson scattering
scross section, and the electron density, integrated over the photon's
path:

$$
{\rm SZ} \approx \int{ \sigma_T n_e dl}
$$

At the same time the electrons in the hot gas that dominates the 
baryonic matter in galaxy clusters also emits X-Rays, and the overall
X-Ray intensity is proportional to the {\it square} of the electron
density integrated along the line of sight through the cluster:

$$
{ \rm X-Ray}  \approx \int{n_e^2 dl}
$$

Using models of the cluster density profile one can then use the the
differing dependence on $n_e$ in the two integrals above to extract 
the physical path-length through the cluster.  Assuming the radial
extension of the cluster is approximately equal to the extension across
the line of sight one can compare the physical size of the cluster to the
angular size to determine its distance.  Clearly, since this assumption
is only good in a statistical sense, the use of S-Z and X-Ray
observations to determine the Hubble constant cannot be done reliably on
the basis of a single cluster observation, but rather on an ensemble.  

A recent preliminary analysis of several clusters \cite{sz}
yields:  

$$
H_0^{SZ} = 60 \pm 10 \ k s^{-1}Mpc^{-1}
$$

\vskip 0.2in
\noindent{ \bf \it Type 1a SN \ (non-Key Project):}

One of the HST Key Project distance estimators involves the use of Type 
1a SN as standard candles.  As previously emphasized, the Key Project
does not perform direct measurements of Type 1a supernovae but rather
uses data obtained by other gorpus.   When these groups perform an
independent analysis to derive a value for the Hubble constant they
arrive at
 a smaller value than that quoted by the Key Project.  Their most recent
quoted value is \cite{1a}:

$$
H_0^{1a} = 64  ^{+8} _{-6} \ k s^{-1}Mpc^{-1}
$$

At the same time, Sandage and collaborators have performed an independent
analysis of SNe Ia distances and obtain \cite{sandage}:

$$
H_0^{1a} = 58  \pm {6} \ k s^{-1}Mpc^{-1}
$$
\vskip 0.2in
\noindent{ \bf \it Surface Brightness Fluctuations and The Galaxy
Density Field:}

Another recently used distance estimator involves the measurement of
fluctuations in the galaxy surface brightness, which correspond to
density fluctuations allowing an estimate of the physical size of a
galaxy.  This measure yields a slightly higher value for the Hubble
constant \cite{davis}:

$$
H_0^{SBF} = 74  \pm 4 \ k s^{-1}Mpc^{-1}
$$

\vskip 0.2in
\noindent{ \bf \it Time Delays in Gravitational Lensing:}

One of the most remarkable observations associated with observations of
multiple images of distant quasars due to gravitational lensing
intervening galaxies has been the measurement of the time delay in the
two images of quasar $Q0957 + 561$.  This time delay, measured quite
accurately to be $ 417 \pm 3$ days is due to two factors:  The
path-length difference between the quasar and the earth for the light
from the two different images, and the Shapiro gravitational time delay
for the light rays  traveling in slightly different
gravitational potential wells.  If it were not for this second factor, a
measurement of the time delay could  be directly used to determine the
distance of the intervening galaxy.  This latter factor however, 
implies that a model of both the galaxy, and the cluster in which it is
embedded must be used to estimate the Shapiro time delay.  This
introduces an additional model-dependent uncertainty into the analysis. 
Two different analyses yield values \cite{chae}:

$$
H_0^{TD1} = 69  ^{+18} _{-12} (1-\kappa) \ k s^{-1}Mpc^{-1}
$$

$$
H_0^{TD2} = 74  ^{+18} _{-10} (1-\kappa) \ k s^{-1}Mpc^{-1}
$$
where $\kappa$ is a parameter which accounts for a possible deviation in 
 cluster parameters governing the overall induced gravitational time delay
of the two signals from that assumed in the best fit.  It is assumed in
the analysis that $\kappa$ is small.

\vskip 0.2in
\noindent{ \bf \it Summary:}

It is difficult to know how to best incorporate all of the quoted
estimates into a single estimate, given their separate systematic and
statistical uncertainties.  Assuming large number statistics, where large
here includes the nine quoted values, I perform a simple weighted average
of the individual estimates, and find an approximate average value:

\begin{equation}
H_0^{Av} \approx 68  \pm 3 \ k s^{-1}Mpc^{-1}
\end{equation}

\subsection{Geometry:}

It has remained a dream of observational cosmologists to be able to
directly measure the geometry of space-time rather than infer the
curvature of the universe by comparing the expansion rate to the mean
mass density.   While several such tests, based on measuring galaxy counts
as a function of redshift, or the variation of angular diameter
distance with redshift, have been attempted in the past, these have all
been stymied by the achilles heel of many observational measurements in
cosmology, evolutionary effects.

Recently, however, measurements of the cosmic microwave background have
finally brought us to the threshold of a direct measurement of geometry,
independent of traditional astrophysical uncertainties.   The idea behind
this measurement is, in principle, quite simple.  The CMB originates from
a spherical shell located at the surface of last scattering (SLS), at a
redshift of roughly $z \approx 1000)$:

\begin{figure}
  \leavevmode\center{\epsfig{figure=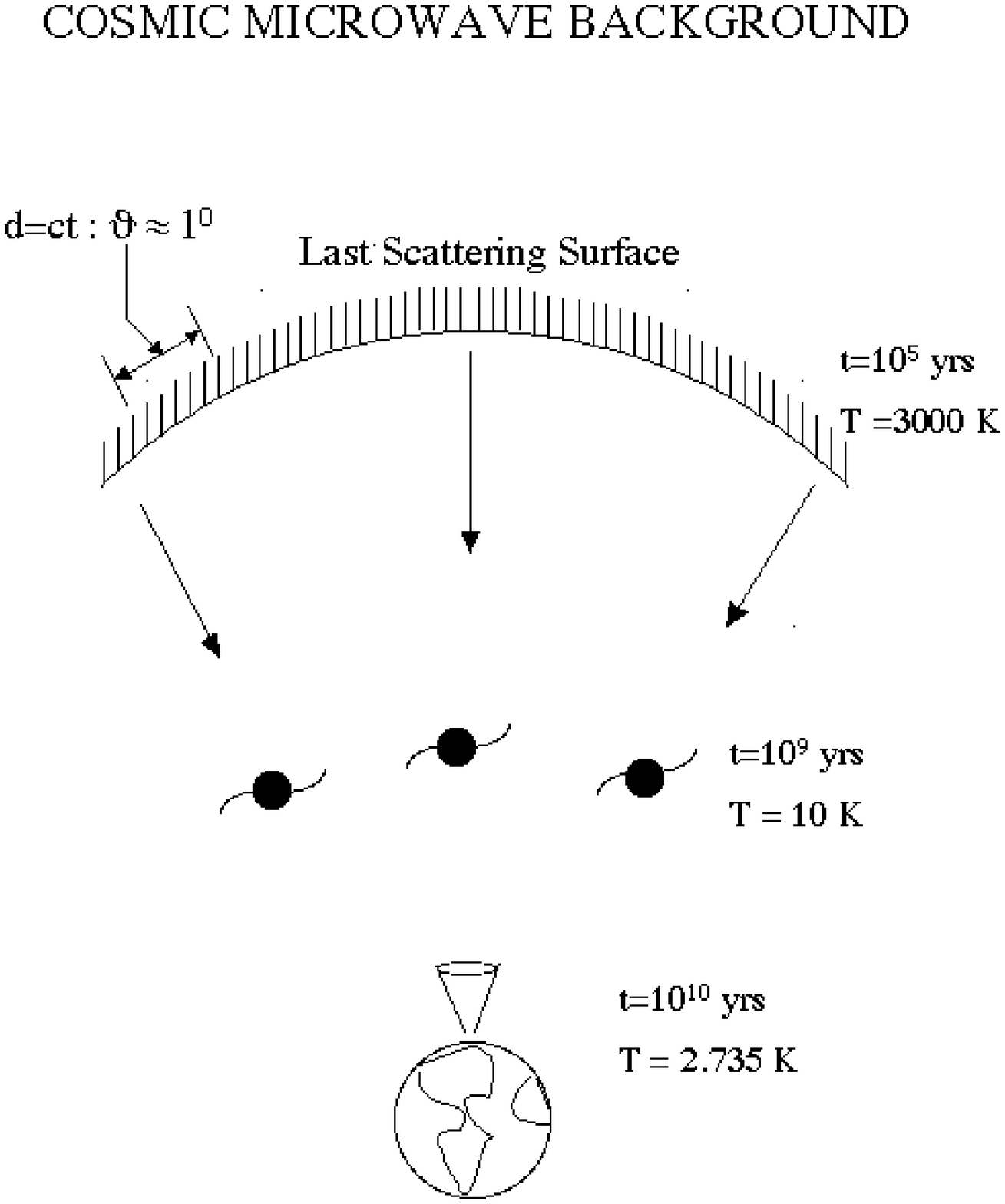,width=8cm}}
\caption{A schematic diagram of the surface of last scattering, showing
the distance traversed by CMB radiation.}
\label{fig:cmb1}
\end{figure}

If a fiducial length could unambigously be distinguished on this surface,
then a determination of the angular size associated with this length
would allow a determination of the intervening geometry:

\begin{figure}
  \leavevmode\center{\epsfig{figure=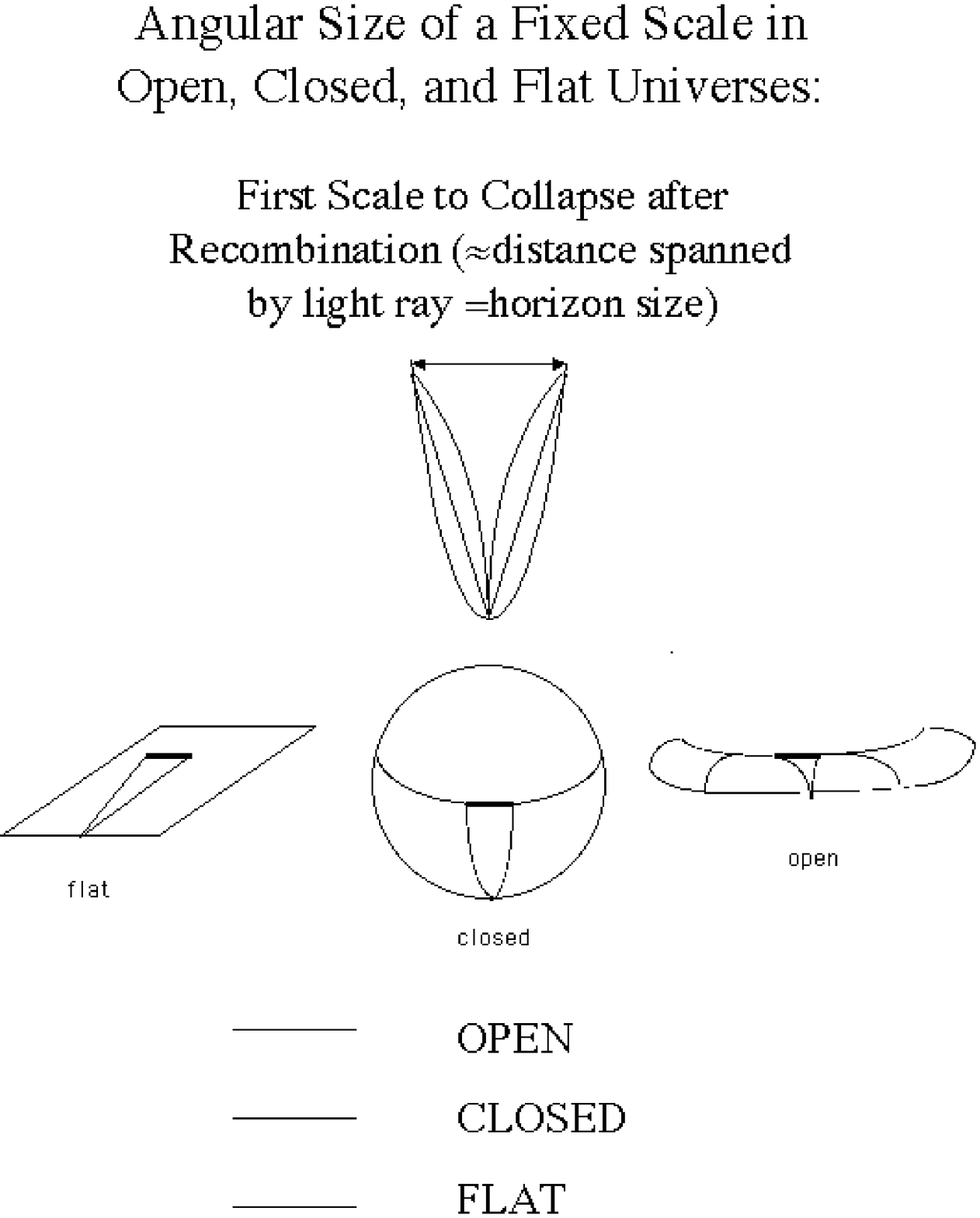,width=8cm}}
\caption{The geometry of the Universe and ray trajectories for CMB
radiation.}
\label{fig:cmb2}
\end{figure}

Fortunately, nature has provided such a fiducial length, which
corresponds roughly to the horizon size at the time the surface of last
scattering existed.  The reason for this is also straightforward.  This
is the largest scale over which causal effects at the time of the
creation of the surface of last scattering could have left an imprint. 
Density fluctuations on such scales would result in acoustic oscillations
of the matter-radiation fluid, and the doppler motion of electrons moving
along with this fluid which scatter on photons emerging from the SLS
produces a characteristic peak in the power spectrum of fluctuations of
the CMBR at a wavenumber corresponding to the angular scale spanned by
this physical scale.  These fluctuations should also be visually
distinguishable in an image map of the CMB, provided a resolution on
degree scales is possible.

Recently, two different ground-based balloon experiments, one launched in
Texas and one launched in Antarctica have resulted in maps with the
required resolution \cite{boomerang,maxima}.  Shown below is a comparison
of the actual boomerang map with several simulations based on a gaussian
random spectrum of density fluctuations in a cold-dark matter universe,
for open, closed, and flat cosmologies.  Even at this qualitative level,
it is clear that a flat universe provides better agreement to between the
simulations and the data than either an open or closed universe
\cite{boom}.   

\begin{figure}
  \leavevmode\center{\epsfig{figure=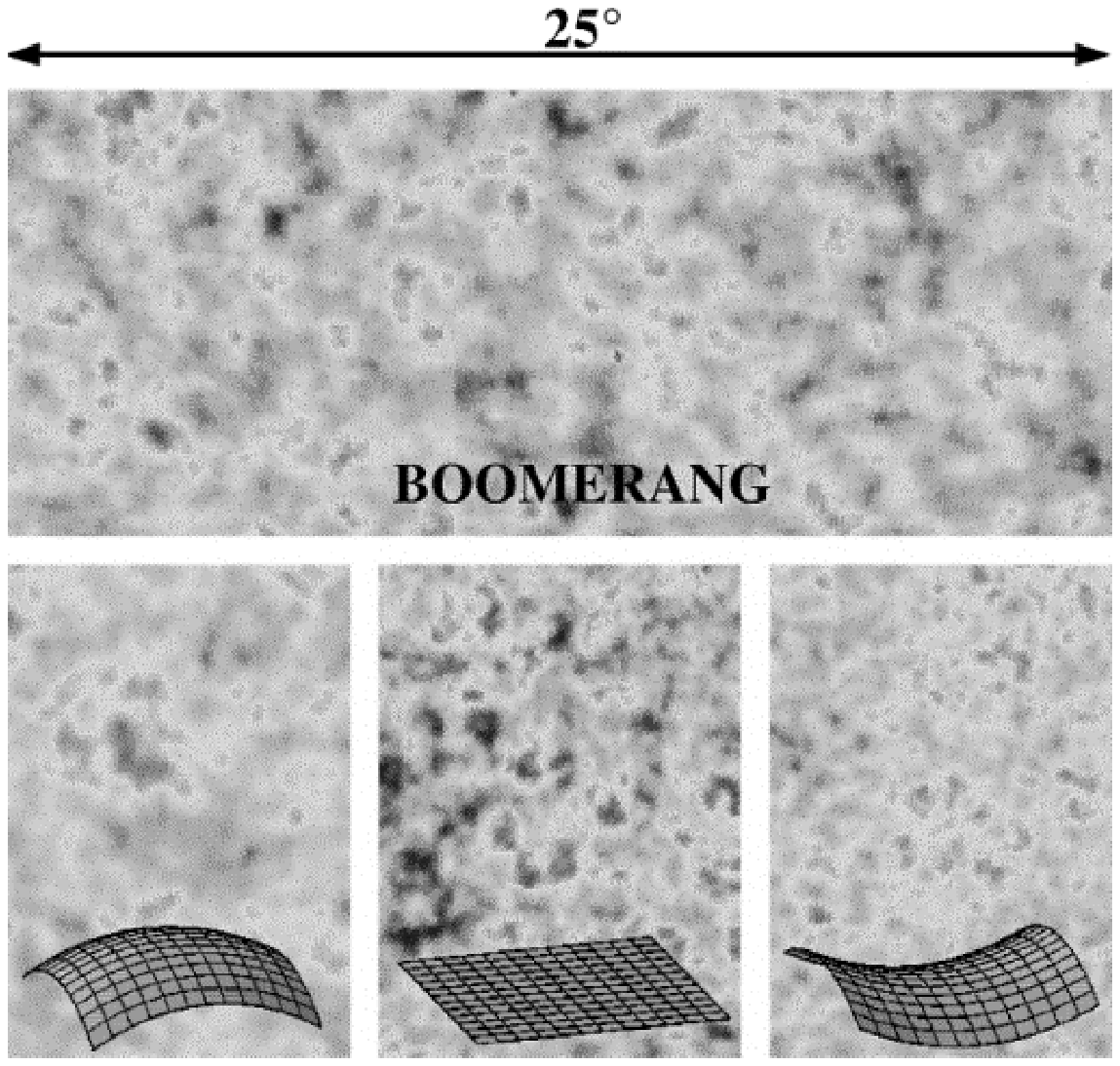,width=10cm}}
\caption{Boomerang data visually compared to expectations for
an open, closed, and flat CDM Universe.}
\label{fig:cmb3}
\end{figure}

On a more quantitative level, one can compare the inferred power spectra
with predicted spectra \cite{teg}.
Such comparisions, for both the Boomerang and Maxima results yields a
constraint on the density parameter:

\begin{equation}
 \Omega = 1.1 \pm .12 (95 \% CL)
\end{equation}

For the first time, it appears that the longstanding prejudice of
theorists, namely that we live in a flat universe, may have been
vindicated by observation!   However, theorists can not be too
self-satisfied by this result, because the source of this energy density
appears to be completely unexpected, and largely inexplicable at the
present time, as we will shortly see.

\section{Time}

\subsection{Stellar Ages:}

Ever since Kelvin and Helmholtz first estimated the age of the Sun to be
less than 100 million years, assuming that gravitational contraction was
its prime energy source, there has been a tension between stellar age
estimates and estimates of the age of the universe.   In the case of the
Kelvin-Helmholtz case, the age of the sun appeared too short to
accomodate an Earth which was several billion years old.  Over much of
the latter half of the 20th century, the opposite problem dominated the
cosmological landscape.    Stellar ages, based on nuclear reactions as
measured in the laboratory, appeared to be too old to accomodate even an
open universe, based on estimates of the Hubble parameter.  Again, as I
shall outline in the next section, the observed expansion rate gives an
upper limit on the age of the Universe which depends upon the equation of
state, and the overall energy density of the dominant matter in the
Universe.  

There are several methods to attempt to determine stellar ages, but I
will concentrate here on main sequence fitting techiniques, because those
are the ones I have been involved in.

The basic idea behind main sequence fitting is simple.  A stellar model is
constructed by solving the basic equations of stellar structure, including
conservation of mass and energy and the assumption of hydrostatic equilibrium,
and the equations of energy transport.  Boundary conditions at the center
of the star and at the surface are then used, and combined with assumed
equation of state equations, opacities, and nuclear reaction rates in
order to evolve a star of given mass, and elemental composition.

Globular clusters are compact stellar systems containing up to $10^5$ stars,
with low heavy element abundance.  Many are located in a spherical halo around
the galactic center, suggesting they formed early in the history of our
galaxy.  By making a cut on those clusters with large halo velocities, and
lowest metallicities (less than 1/100th the solar value), one attempts to
observationally distinguish the oldest such systems. Because these systems are
compact, one can safely assume that all the stars within them formed at
approximately the same time.

Observers measure the color and luminosity of stars in such clusters, producing
color-magnitude diagrams of the type shown in Figure 2 (based on data 
from \cite{durr}.

\begin{figure}
  \leavevmode\center{\epsfig{figure=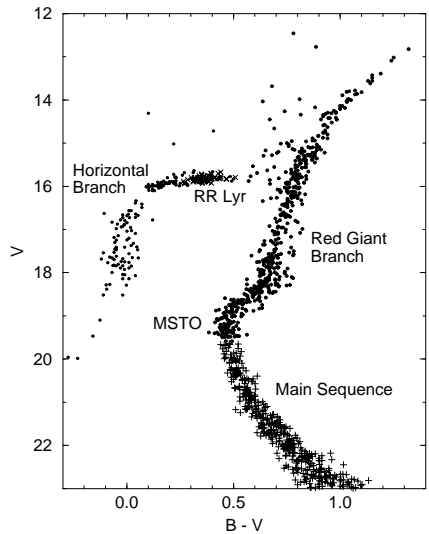,width=8cm}}
\caption{Color-magnitude diagram for a typical globular
cluster, M15. Vertical axis plots the magnitude (luminosity) of
the stars in the V wavelength region and the horizontal axis plots the
color (surface temperature) of the stars.}
\label{fig:CMDIAG}
\end{figure}

Next, using stellar models, one can attempt to evolve stars of differing
mass for the metallicities appropriate to a given cluster, in order to fit
observations.  A point which is often conveniently chosen is the so-called main
sequence-turnoff (MSTO) point, the point in which hydrogen burning (main
sequence) stars have exhausted their supply of hydrogen in the core.  After the
MSTO, the stars quickly expand, become brighter, and are referred to as Red
Giant Branch (RGB) stars.  Higher mass stars develop a helium core that is so
hot and dense that helium fusion begins.  These form along the horizontal
branch.  Some stars along this branch are unstable to radial pulsations, the
so-called RR Lyrae stars mentioned earlier, which are important distance
indicators.  While one in principle could attempt to fit theoretical isochrones
(the locus of points on the predicted CM curve corresponding to different mass
stars which have evolved to a specified age), to observations at any point, the
main sequence turnoff is both sensitive to age, and involves minimal
(though just how minimal remains to be seen) theoretical uncertainties.

Dimensional analysis tells us that the main sequence turnoff should be a
sensitive function of age.  The luminosity of main sequence stars is very
roughly proportional to the third power of solar mass.  Hence the time it takes
to burn the hydrogen fuel is proportional to the total amount of fuel
(proportional to the mass M), divided by the Luminosity---
proportional to $M^3$.  Hence the lifetime of stars on the main sequence
is roughly proportional to the inverse square of the stellar mass.

Of course the ability to go beyond this rough approximation depends completely
on the on the confidence one has in one's stellar models.  It is worth
noting that several improvements in stellar modeling have recently
combined to lower the overall age estimates of globular clusters.  The
inclusion of diffusion lowers the age of globular clusters by about 7$\%$
\cite{eighteen}, and a recently improved equation of state which incorporates
the effect of Coulomb interactions \cite{nineteen} has lead to a further
7$\%$ reduction in overall ages.   Of course, what is most important for
the comparison of cosmological predictions with inferred age estimates is
the uncertainties in these and other stellar model parameters, and
not merely their best fit values. 

Over the course of the past several years, I and my collaborators have
tried to incorporate stellar model uncertainties, along with
observational uncertainties into a self consistent Monte Carlo analysis
which might allow one to estimate a reliable range of globular cluster
ages.   Others have carried out independent, but similar studies, and at
the present time, rough agreement has been obtained between the different
groups (i.e. see\cite{krauss}).  

I will not belabor the detailed history of all such efforts here.  The
most crucial insight has been that stellar model uncertainties
are small in comparison to an overall observational uncertainty inherent
in fitting predicted main sequence luminosities to observed turnoff
magnitudes.  This matching depends crucially on a determination of the
distance to globular clusters.  The uncertainty in this distance scale
produces by far the largest uncertainty in the quoted age estimates.

In many studies, the distance to globular clusters can be parametrized in
terms of the inferred magnitude of the horizontal branch stars.  This
magnitude can, in turn, be presented in terms of the inferred absolute
magnitude,
$M_v(\rm RR)$of RR Lyrae
variable stars located on the horizontal branch. 

In 1997, the Hipparcos satellite produced its catalogue of 
parallaxes of nearby stars,
causing an apparent revision in distance estimates.  The Hipparcos parallaxes
seemed to be systematically smaller, for the smallest measured parallaxes, than
previous terrestrially determined parallaxes.  Could this represent the
unanticipated systematic uncertainty that David has suspected?  Since all 
the
detailed analyses had been pre-Hipparcos, several groups scrambled to
incorporate the Hipparcos catalogue into their analyses.  The immediate result
was a generally lower mean age estimate, reducing the mean value to 11.5-12 Gyr,
and allowing ages of the oldest globular clusters as low as 9.5 Gyr.   However,
what is also clear is that there is now an explicit systematic uncertainty in
the RR Lyrae distance modulus which dominates the results.  Different
measurements are no longer consistent.  Depending upon which distance estimator
is correct, and there is now better evidence that the distance estimators which
disagree with Hipparcos-based main sequence fitting should not be dismissed out
of hand, the best-fit globular cluster estimate could shift up perhaps $1
\sigma$, or about 1.5 Gyr, to about 13 Gyr.

Within the past two years, Brian Chaboyer and I have reanalyzed globular
cluster ages, incorporating new nuclear reaction rates, cosmological
estimates of the $^4$He abundance, and most importantly, several new
estimates of $M_v(\rm RR)$.   The result is that while systematic
uncertainties clearly still dominate, we argue that the mean age of the
oldest globular clusters has increased about 1 Gyr, to be 
$12.7^{+3}_{-2} $ ($95\%$) Gyr, with a 95 $\%$ confidence range of about
11-16 Gyr \cite{chabrkau}.   It is this range that I shall now compare to
that determined using the Hubble estimates given earlier.

\subsection{Hubble Age:}

As alluded to earlier, in a Friedman-Robertson-Walker Universe, the age of
the Universe is directly related to both the overall density of energy,
and to the equation of state of the dominant component of this energy
density.  The equation of state is parameterized by the ratio $\omega = p
/{\rho}$, where $p$ stands for pressure and $\rho$ for energy
density.  It is this ratio which enters into the second order Friedman
equation describing the change in Hubble parameter with time, which in
turn determines the age of the Universe for a specific net total energy
density.

 The fact that this depends on two independent parameters has
meant that one could reconcile possible conflicts with globular cluster
age estimates by altering either the energy density, or the equation of
state.   An open universe, for example, is older for a given Hubble
Constant, than is a flat universe, while a flat universe dominated by a
cosmological constant can be older than an open matter dominated
universe.

If, however, we incorporate the recent geometric determination which
suggests we live in a flat Universe into our analysis, then our
constraints on the possible equation of state on the dominant energy 
density of the universe become more severe.   If, for existence, we
allow for a diffuse component to the total energy density with the
equation of state of a cosmological constant ($\omega =-1$), then the
age of the Universe for various combinations of matter and cosmological
constant are shown below.

\begin{table}[h]
\caption{Hubble Ages for a Flat Universe, $H_0 = 68 \pm 6$, ("$2
\sigma$")}
\begin{center}
\begin{tabular}{|l|l| c|}
\hline
 $\Omega_M$ & $\Omega_x$ & $t_0$ \\ \hline
 $1$ & $0$ & $9.7 \pm 1$ \\
 $0.2$ & $0.8$ & $15.3 \pm 1.5$ \\
 $0.3$ & $0.7$ & $13.7 \pm 1.4$ \\
 $0.35$ & $0.65$ & $12.9 \pm 1.3$ \\ \hline
\end{tabular}
\end{center}
\end{table}

Clearly, a matter-dominated flat universe is in trouble if one wants to
reconcile the inferred Hubble age with the lower limit on the age of the
universe inferred from globular clusters.  In fact, if one took the
above constraints at face value, { \it such a Universe is ruled out on
the basis of age estimates and the Hubble constant estimates}.  However, I
am old enough to know that systematic uncertainties in cosmology often
shift parameters well outside their formal two sigma, or even three sigma
limits.  In order to definitely rule out a flat matter dominated universe
using a comparison of stellar and Hubble ages, uncertainties in both
would have to be reduced by at least a factor of two. 

\section{Matter}

Having indirectly probed the nature of matter in the Universe using the
previous estimates, it is now time to turn to direct constraints that
have been derived in the past decade.  Here, perhaps more than any other
area of observational cosmology, new observations have changed the way we
think about the Universe.  

\subsection{The Baryon Density: a re-occuring crisis?:}

The success of Big Bang Nucleosynthesis in predicting in the cosmic
abundances of the light elements has been much heralded.  Nevertheless,
the finer the ability to empirically infer the primordial abundances on
the basis of observations, the greater the ability to uncover some
small deviation from the predictions.   Over the past five years, two
different sets of observations have threatened, at least in some people's
minds, to overturn the simplest BBN model predictions.  I believe it is
fair to say that most people have accepted that the first threat was
overblown.  The concerns about the second have yet to fully subside.
\vskip 0.1in

{\it i. Primordial Deuterium:}  The production of primordial deuterium
during BBN is a monotonically decreasing function of the baryon density
simply because the greater this density the more efficiently protons and
neutrons get processed to helium, and deuterium, as an intermediary
in this reactions set, is thus also more efficiently processed at the
same time.   The problem with inferring the primordial deuterium
abundance by using present day measurements of deuterium abundances in
the solar system, for example, is that deuterium is highly processed
(i.e. destroyed) in stars, and no one has a good enough model for
galactic chemical evolution to work backwards from the observed
abundances in order to adequately constrain deuterium at a level where
this constraint could significantly test BBN estimates.

Three years ago, the situation regarding deuterium as a probe of BBN
changed dramatically, when David Tytler and Scott Burles convincingly
measured the deuterium fraction in high redshift hydrogen clouds that
absorb light from even higher redshift quasars.   Because these clouds
are at high redshift, before significant star formation has occurred,
little post BBN deuterium processing is thought to have taken place, and
thus the measured value gives a reasonable handle on the primordial BBN
abundance.  The best measured system \cite{tytler} yields a deuterium to
hydrogen fraction of 

\begin{equation}
(D/H) = (3.3. \pm 0.5) \times 10^{-5} \ \  ( 2 \sigma )
\end{equation}

This, in turn, leads to a contraint on the baryon fraction of the
Universe, via standard BBN, 

\begin{equation}
\Omega_{B} h^2 = .0190 \pm .0018 \ \  ( 2 \sigma )
\end{equation}

where the quoted uncertainty is dominated by the observational
uncertainty in the D/H ratio, and where $H_0 =100 h$.  Thus, taken at face
value, we now know the baryon density in the universe today to an
accuracy of about
$10 \%$!

When first quoted, this result sent shock waves through some of the
BBN community, because this value of $\Omega_B$ is only consistent if the
primordial helium fraction (by mass) is greater than about 24.5$\%$. 
However, a number of previous studies had claimed an upper limit well
below this value.  After the dust has settled, it is clear that
these previous claims are likely to under-estimated systematic
observational effects.  Recent studies, for example, place an upper limit
on the primordial helium fraction closer to 25$\%$.  

In any case, even if somehow the deuterium estimate is wrong, one can
combine all the other light element constraints to produce a range
for $\Omega_b h^2$ consistent with observation:

\begin{equation}
\Omega_{B} h^2 = .016 -0.025
\end{equation}

\vskip 0.1in
{\it ii. CMB constraints:} Beyond the great excitement over the
observation of a peak in the CMB power spectrum at an angular scale
corresponding to that expected for a flat universe lay some
excitement/concern over the small apparent size of the next peak in the
spectrum, at higher multipole moment (smaller angular size).  The height
of the first peak in the CMB spectrum is related to a number of
cosmological parameters and thus cannot alone be used to constrain any
one of them.  However, the relative height of the first and second peaks
is strongly dependent on the baryon fraction of the universe, since the
peaks themselves arise from compton scattering of photons off of
electrons in the process of becoming bound to baryons.   Analyses of the
two most recent small-scale CMB results produces a claimed constraint
\cite{teg}:

\begin{equation}
\Omega_{B} h^2 = .032\pm .009 \ \  ( 2 \sigma )
\end{equation}
  
Depending upon how you look at this, this is either a stunning
confirmation that the overall scale for $\Omega_B$ predicted by simple
BBN analyses is correct, or a horrible crisis, in which the two
constraints, one from primordial deuterium, and one from CMB
observations, disagree at the two sigma level.   Given the history of
this subject, I expect the former response is perhaps appropriate for the
moment.  In particular, the CMB results are the very first to probe this
regime, and first observations are often suspect, and in addition, the
CMB peak heights do have a dependence on other cosmological parameters
which must be fixed in order to derive the above constraint on
$\Omega_B$. 

Assuming  the Burles and Tytler limit on $\Omega_B h^2$ is correct, and
taking the range for $H_0$ given earlier, one derives the constraint on
$\Omega_B$ of

\begin{equation}
\Omega_{B} = .045 \pm 0.15
\end{equation}

\subsection{ $\Omega_{matter}$}

Perhaps the greatest change in cosmological prejudice in the past decade
relates to the inferred total abundance of matter in the Universe. 
Because of the great intellectual attraction Inflation as a mechanism to
solve the so-called Horizon and Flatness problems in the Universe, it is
fair to say that most cosmologists, and essentially all particle
theorists had implicitly assumed that the Universe is flat, and thus that
the density of dark matter around galaxies and clusters of galaxies was
sufficient to yield $\Omega =1$.   Over the past decade it became more
and more difficult to defend this viewpoint against an increasing number
of observations that suggested this was not, in fact, the case in the
Universe in which we live.

The earliest holes in this picture arose from measurements of galaxy
clustering on large scales.  The transition from a radiation to matter
dominated universe at early times is dependent, of course, on the total
abundance of matter.  This transition produces a characteristic signature
in the spectrum of remnant density fluctuations observed on large
scales.  Making the assumption that dark matter dominates on large
scales, and moreover that the dark matter is cold (i.e. became
non-relativistic when the temperature of the Universe was less than about
a keV), fits to the two point correlation function of galaxies on large
scales yielded \cite{peac,liddle}:

\begin{equation}
\Omega_M h =.2-.3
\end{equation}

Unless $h$ was absurdly small, this would imply that $\Omega_M$ is
substantially less than 1.

The second nail in the coffin arose when observations of the evolution
of large scale structure as a function of redshift began to be made.
Bahcall and collaborators \cite{neta} argued strongly that evidence for
any large clusters at high redshift would argue strongly against a flat
cold dark matter dominated universe, because in such a universe structure
continues to evolve with redshift up to the present time on large scales,
so that in order to be consistent with the observed structures at low
redshift, far less structure should be observed at high redshift.  Claims
were made that an upper limit $\Omega_B \le 0.5$ could be obtained by
such analyses.

A number of authors have questioned the systematics inherent in the early
claims, but it is certainly clear that there appears to be more structure
at high redshift than one would naively expect in a flat matter dominated
universe.  Future studies of X-ray clusters, and use of the
Sunyaev-Zeldovich effect to measure cluster properties should be able to
yield measurements which will allow a fine-scale distinction not just
between models with different overall dark matter densities, but also
models with the same overall value of $\Omega$ and different equations
of state for the dominant energy \cite{mohr}.

For the moment, however, perhaps the best overall constraint on the total
density of clustered matter in the universe comes from the combination of
X-Ray measurements of clusters with large hydrodynamic simulations.  The
idea is straightforward.  A measurement of both the temperature and
luminosity of the X-Rays coming from hot gas which dominates the total
baryon fraction in clusters can be inverted, under the assumption of
hydrostatic equilibrium of the gas in clusters, to obtain the underlying
gravitational potential of these systems.  In particular the ratio of
baryon to total mass of these systems can be derived.   Employing the
constraint on the total baryon density of the Universe coming from BBN,
and assuming that galaxy clusters provide a good mean estimate of the
total clustered mass in the Universe, one can then arrive at an allowed
range for the total mass density in the Universe
\cite{white,krauss2,evrard}.  Many of the initial systematic
uncertainties in this analysis having to do with cluster modelling have
now been dealt with by better observations, and better simulations
( i.e. see\cite{mohr2}), so that now a combination of BBN and cluster
measurements yields:
\begin{equation}
\Omega_M = 0.35 \pm 0.1  \ \ (2\sigma)
\end{equation}

\subsection{Equation of State of Dominant Energy:}

Remarkably, the above estimate for $\Omega_M$ brings the
discussion of cosmological parameters full circle, with consistency
obtained for a flat 12.5 billion year old universe , but not one dominated
by matter.  As noted previously, a cosmological constant dominated 
universe with $\Omega_M = 0.35$ has an age which nicely fits in the
best-fit range.   However, based on the data discussed thus far, we have
no direct evidence that the dark energy necessary to result in a flat
universe actually has the equation of state appropriate for a vacuum
energy.  Direct motivation for the possibility that the dominant energy
driving the expansion of the Universe violates the Strong Energy
Condition came, in 1998, from two different sets of observations of
distant Type 1a Supernovae.  In measuring the distance-redshift relation
\cite{perl,kirsh} these groups both came to the same, surprising
conclusion:  the expansion of the Universe seems to be accelerating! 
This is only possible if the dominant energy is
"cosmological-constant-like", namely if "$\omega < -0.5$  (recall that
$\omega =-1$ for a cosmological constant).  

In order to try and determine if the dominant dark energy does in fact
differ significantly from a static vacuum energy---as for example may
occur if some background field that is dynamically evolving is
dominating the expansion energy at the moment---one can hope to search for
deviations from the distance-redshift relation for a cosmological
constant-dominated universe.  To date, none have been observed.  In
fact, existing measurements already put an upper limit $\omega \le -0.6$
\cite{perlturn}.   Recent work \cite{me} suggests that the best one
might be able to do from the ground using SN measurements would be to
improve this limit to $\omega \le -0.7$.  Either other measurements,
such as galaxy cluster evolution observations, or space-based SN
observations would be required to further tighten the constraint.

\section{Conclusions: A Cosmic Uncertainty Principle}

I list the overall constraints on cosmological parameters discussed in
this review in the table below.  It is worth stressing how completely
remarkable the present situation is.  After 20 years, we now have the
first direct evidence that the Universe might be flat, but we also have
definitive evidence that there is not enough matter, including dark
matter, to make it so.  We seem to be forced to accept the possibility
that some weird form of dark energy is the dominant stuff in the
Universe.  It is fair to say that this situation is more mysterious, and
thus more exciting, than anyone had a right to expect it to be.

\begin{table}[h]
\caption{Cosmological Parameters 2001}
\begin{center}
\begin{tabular}{|l|l| c|}
\hline
Parameter & Allowed range  & Formal Conf. Level (where approp.) \\
\hline
 $H_0$ & $68 \pm 6$ & $ 2 \sigma$ \\
 $t_0$ & $12.7^{+3}_{-2}$ & $2 \sigma$ \\
 $\Omega_B h^2$ & $.019 \pm .0018$ or $.032 \pm .009$ & \\
 $\Omega_B$ & $0.045 \pm 0.015 $ & $2 \sigma $ \\ 
 $\Omega_M$ & $0.35 \pm 0.1 $ & $2 \sigma $ \\
 $\Omega_{TOT}$ & $1.1 \pm 0.12 $ & $2 \sigma $ \\
 $\Omega_{X}$ & $0.65 \pm 0.15 $ & $2 \sigma $ \\
 $\omega$ & $\le -0.6$ & $2 \sigma $ \\ \hline
\end{tabular}
\end{center}
\end{table}

The new situation changes everything about the way we think about
cosmology.  In the first place, it demonstrates that Geometry and Destiny
are no longer linked.  Previously, the holy grail of cosmology involved
determining the density parameter $\Omega$, because this was tantamount
to determining the ultimate future of our universe.  Now, once we accept
the possibility of a non-zero cosmological constant, we must also accept
the fact that any universe, open, closed, or flat, can either expand
forever, or reverse the present expansion and end in a big crunch
\cite{kraussturn}.  But wait, it gets worse, as my colleague Michael
Turner and I have also demonstrated, there is no set of cosmological
measurements, no matter how precise, that will allow us to determine the
ultimate future of the Universe.  In order to do so, we would require a
theory of everything.   

On the other hand, if our universe is in fact dominated by a cosmological
constant, the future for life is rather bleak \cite{kraussstark}. 
Distant galaxies will soon blink out of sight, and the Universe will
become cold and dark, and uninhabitable....   

This bleak picture may seem depressing, but the flip side of all the
above is that we live in exciting times now, when mysteries abound.  And
these mysteries, or new ones, are likely to remain with us for some time.

\section*{Acknowledgments}

I thank my collaborators involved in various aspects of my own work
described here, including Michael Turner, Brian Chaboyer, 
Craig Copi, and Glenn Starkman, and also the observers whose results
have helped make cosmology so exciting in the past decade.
I also thank the the organizers for a very stimulating and enjoyable 
meeting.

\end{document}